\documentclass[aip, prb, preprint]{revtex4-2}

\usepackage{amsmath}    % need for subequations
\usepackage{graphicx}   % need for figures
\usepackage{verbatim}   % useful for program listings
\usepackage{color}      % use if color is used in text
\usepackage{epstopdf}   % use for side-by-side figures
\usepackage{upgreek}    % for non-italic greek letters (units)
\usepackage[hidelinks]{hyperref}   % use for hypertext links, including those to external documents and URLs
\usepackage{textcomp}   % Trademark symbol
\usepackage[separate-uncertainty=true, multi-part-units=single]{siunitx}
\usepackage[capitalise]{cleveref}

\begin{document}

\title{Influence of magnetic fields on ultrafast laser-induced switching dynamics in Co/Gd bilayers}

\author{M.J.G. Peeters}
\email[E-mail: ]{m.j.g.peeters@tue.nl}
\affiliation{Department of Applied Physics, Eindhoven University of Technology, PO Box 513, 5600 MB Eindhoven, The Netherlands}

\author{Y.M. van Ballegooie}
\affiliation{Department of Applied Physics, Eindhoven University of Technology, PO Box 513, 5600 MB Eindhoven, The Netherlands}

\author{B. Koopmans}
\affiliation{Department of Applied Physics, Eindhoven University of Technology, PO Box 513, 5600 MB Eindhoven, The Netherlands}

\date{\today}

\begin{abstract}
Recently it has been shown that not only GdFeCo alloys exhibit single-pulse helicity-independent all-optical switching (HI-AOS), but that this effect is also seen in Co/Gd bilayers. However, there have been no reports on the explicit time dynamics of the switching process in these bilayers as of yet. Furthermore, time-resolved measurements of switching of other materials are typically done with a constant applied field to reset the magnetization between consecutive pulses and thus ensure repeatable behavior. In this paper we experimentally resolve the explicit dynamics of the switching process in Co/Gd, and the influence of applied magnetic fields on the switching process. We observe that after a switch within several picoseconds, the magnetization switches back at a timescale of hundreds of picoseconds. This backswitch includes a strong dependence on the magnetic field strength even at sub-tesla fields, significantly smaller than the exchange fields that govern the switching dynamics. This surprising behaviour is explained by a combination of longitudinal switching (on a picosecond timescale), precessional switching (on a nanosecond timescale) and domain-wall motion (on a timescale of \SI{10}{\nano\second} and beyond). We discuss these different switching regimes and their relative importance using simple model calculations.
\end{abstract}

\maketitle

\section{Introduction}
\label{sec:introduction}
All-optical switching (AOS) of the magnetization is a phenomenon where the direction of the magnetization in a material can be switched using laser pulses \cite{Stanciu_PRL_2007, ElHadri_PRB_2016, Hohlfeld_APL_2009, Steil_PRB_2011, Lambert_Science_2014, Mangin_NatMat_2014}. It has the potential to be used in new types of data storage devices, since optically writing magnetic bits provides unprecedented speeds and high energy efficiency. First discovered in 2007, single-pulse helicity-independent all-optical switching (HI-AOS) was originally only found in GdFeCo alloys\cite{Stanciu_PRL_2007}. It is generally believed that the anti-ferromagnetic coupling between the two spin sublattices (Gd and FeCo) is essential for AOS to appear. Upon excitation with a laser pulse, both subsystems demagnetize on different timescales \cite{Radu_Nature_2011}. The transfer of angular momentum between the two sublattices, mediated by exchange scattering, is thought to be the driving force behind the switching\cite{Ostler_NatComms_2012}. Recently, it has been shown that AOS is also possible in Co/Gd bilayers, which is extremely interesting from an application point of view as it provides opportunities for interface engineering\cite{Lalieu_PRB_2017, Lalieu_NatComms_2019}. 

However, there have been no reports on the explicit time-resolved dynamics of the switching process in Co/Gd bilayers, raising the question whether switching in Co/Gd bilayers happens on the same ultrafast \si{\pico\second}-timescale as with the CoGd and GdFeCo alloys. Furthermore, most experiments on the dynamics of the switching process in AOS with time-resolved MOKE (TR-MOKE) or similar techniques have been performed using a constant applied magnetic field \cite{Radu_Nature_2011,ElGhazaly_APL_2019}. This field is used to return the magnetization to its original direction between successive laser pulses. If the applied magnetic field has an influence on the measured dynamics, it is important to keep this in mind when analyzing time-resolved AOS experiments. Consequently, we want to answer a second question; how big is the influence of this field on the dynamics, and on what timescales does this influence become of relevance.

In this paper we will show a TR-MOKE method that can be used to measure the switching dynamics without applying a magnetic field. In Co/Gd (and GdFeCo) the magnetization switches via a highly stable toggle mechanism, alternating between up and down. A constant applied field can be used to return the magnetization to its original direction between successive laser pulses, but in our method we detect every other laser pulse, thereby measuring either the up-down or the down-up switch without the need for an applied magnetic field. This is confirmed by the demonstration of switching for at least $10^8$ pulses at a repetition rate of \SI{100}{\kilo\hertz}. In these time-resolved measurements we observe a similar ultrafast switching response in Co/Gd bilayers as in GdFeCo alloys. After that, we report on the field-dependence of the switching process, where a constant field of \SIrange[range-units = single]{0}{400}{\milli\tesla} is applied in the initial direction of the magnetization. We find that the initial switch and recovery to about 50 \% of the saturation magnetization (in the opposite direction) occurs in approximately \SIrange{10}{20}{\pico\second} and is independent of the applied field. However, from that moment onwards the dynamics strongly depend on the applied field, and a field of several \SI{100}{\milli\tesla} suffices to obtain a backswitch within a few \SI{100}{\pico\second}. This is remarkable, as the exchange fields that govern the switching process are orders of magnitude higher than the applied fields that we use. We will discuss the various mechanisms that can play a role in this backswitch, namely longitudinal switching (described by a layered-M3TM \cite{Beens_PRB_2019}), precessional switching (described by an LLG model) and switching via domain-wall motion, and we will examine the importance of these processes in the backswitch. We conclude that the backswitch is dominated by precessional switching, although not all features of the switching process can be explained by precessional switching alone.

\section{Methods}
\label{sec:methods}
The measurements are performed on SiB//Ta (\SI{4}{\nano\meter})/Pt (\SI{4}{\nano\meter})/Co (\SI{1}{\nano\meter})/Gd (\SI{3}{\nano\meter})/Pt (\SI{4}{\nano\meter}) samples, which are grown by DC magnetron sputter deposition at room temperature, with a base pressure of $10^{-8} - 10^{-9}$\si{\milli\bar}. The Co and Gd are coupled anti-ferromagnetically, with roughly \SI{0.5}{\nano\meter} of the Gd magnetized\cite{Lalieu_PRB_2017}. This proximity-induced Gd magnetization is in the opposite direction of the Co magnetization. Hysteresis loops show square loops\cite{Lalieu_PRB_2017}, confirming a perpendicular magnetic anisotropy, with coercive fields of about \SI{20}{\milli\tesla}. The magneto-optic signal in our experiments is a linear combination of contributions by the Co and Gd subsystems, but is generally assumed to be dominated by Co in our wavelength range \cite{Khorsand_PRL_2013, Arenholz_PRB_1995, Lang_1981}.

The time-resolved experiments are carried out with a Spectra-Physics Spirit-NOPA laser system. The pulse length at sample position is $\sim$\SI{100}{\femto\second}, with a central wavelength of \SI{700}{\nano\meter} at a repetition rate of \SI{100}{\kilo\hertz}. A high-aperture laser objective is used to focus the pump and probe pulses on the sample to a size (full width at half maximum) of $\sim$\SI{80}{\micro\meter} for the pump pulses and $\sim$\SI{30}{\micro\meter} for the probe pulses. In order to measure the switching dynamics without an applied field we want to measure only the even or odd pulses. This is achieved by adding a transistor between the photo-detector and the lock-in amplifier. By synchronizing the signal applied to the base of the transistor with the laser we are able to block exactly half of the pulses in the switching process. A more detailed explanation about the setup can be found in Supplementary Note I \cite{supplementary}.

In order to investigate the robustness of the switching process we expose the sample to laser pulses at \SI{125}{\kilo\hertz} for increasing amounts of time. The result can be seen in \cref{fig:figure1CoGd}a, confirming that we achieve robust switching for up to one hour, corresponding to $4.5\cdot 10^8$ pulses. For too high laser fluences the combination of a relatively high repetition rate and the long exposure time causes damage to the sample, either by annealing the stack or by causing structural damage in the stack (not shown here). In our experiments, care is taken to avoid this damage.

\section{Field-free switching}
\begin{figure}[tb]
	\centering
	\includegraphics[width=\columnwidth]{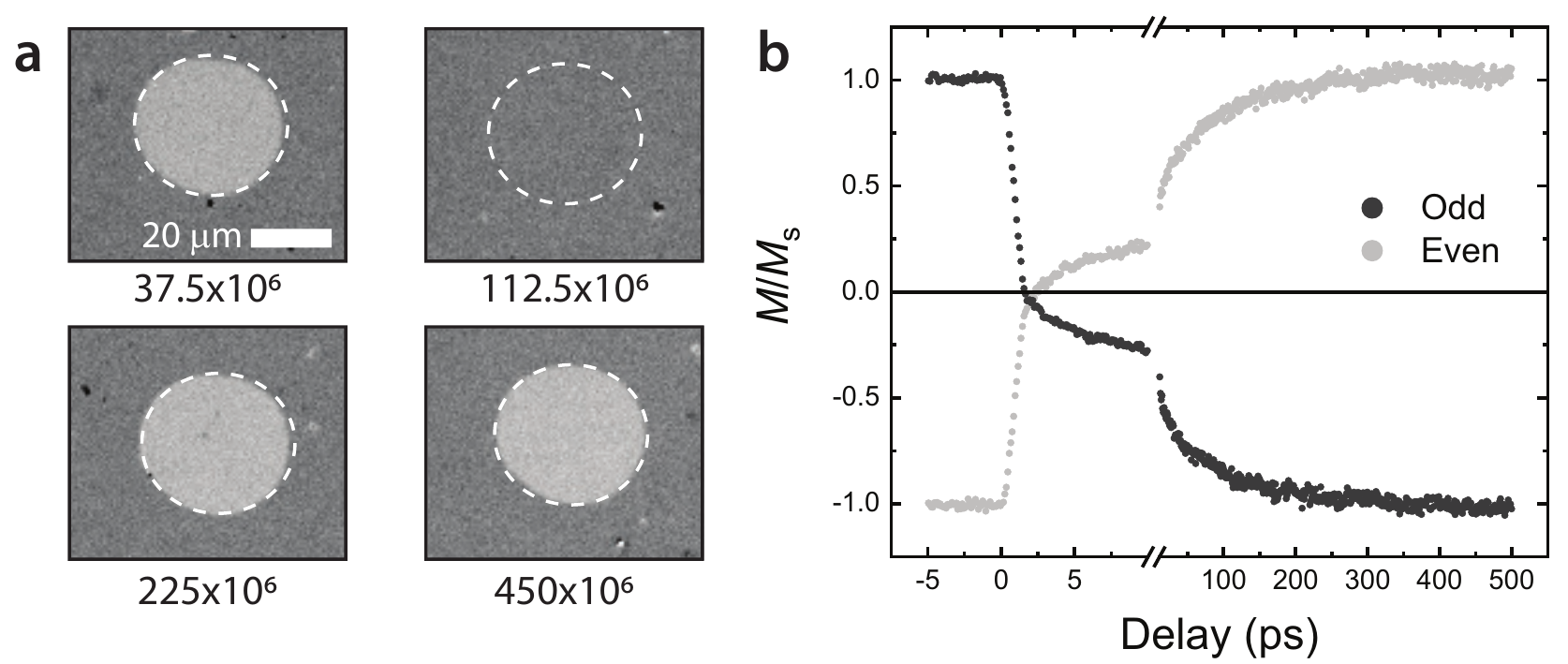}
	\caption{a) Kerr microscope images of AOS in Co/Gd. Dark and light contrast corresponds to up and down magnetization, the dashed line indicates the region where the local fluence is high enough to reverse the magnetization. A switched domain is observed depending on whether the total number of pump pulses is even or odd. The labels indicate the (approximate) number of pulses for each image. The measurements are done with a repetition rate of \SI{125}{\kilo\hertz}. b) Field-free time-resolved measurement of AOS in Co/Gd, by measuring only half the pulses, at \SI{100}{\kilo\hertz}. Using only the odd or even pulses corresponds to measuring the up-down or down-up switch, respectively. Note the axis break on the $x$-axis. }
	\label{fig:figure1CoGd}
\end{figure}

In \cref{fig:figure1CoGd}b we show the TR-MOKE measurement on the switching dynamics in Co/Gd without applying a magnetic field, selecting either even or odd pulses. Time traces for even and odd pulses are exactly opposite, confirming the highly deterministic toggle switching. We are able to switch between measuring the even and odd pulses by changing the phase of the signal sent to the base of the transistor. In Supplementary Note II \cite{supplementary} we confirm that the measurements with transistor do not introduce any artifacts in the measured dynamics. As can be seen left of the axis break in \cref{fig:figure1CoGd}b, within several \si{\pico\second} switching occurs, which is similar to the ultrafast switching seen in GdCo and GdFeCo alloys \cite{ElGhazaly_APL_2019}. This answers our first question, i.e. switching in Co/Gd bilayers occurs at the same timescale as in GdFeCo alloys. On the right side of the axis break, after about \SI{200}{\pico\second} we see a complete recovery of the magnetization in the switched direction due to the cooling down of the system. We note that this recovery is due to the cooling down of the film, governed by the thermal diffusion into the substrate. 

\section{Field-dependent switching}
\begin{figure}[tb]
	\centering
	\includegraphics[width=\columnwidth]{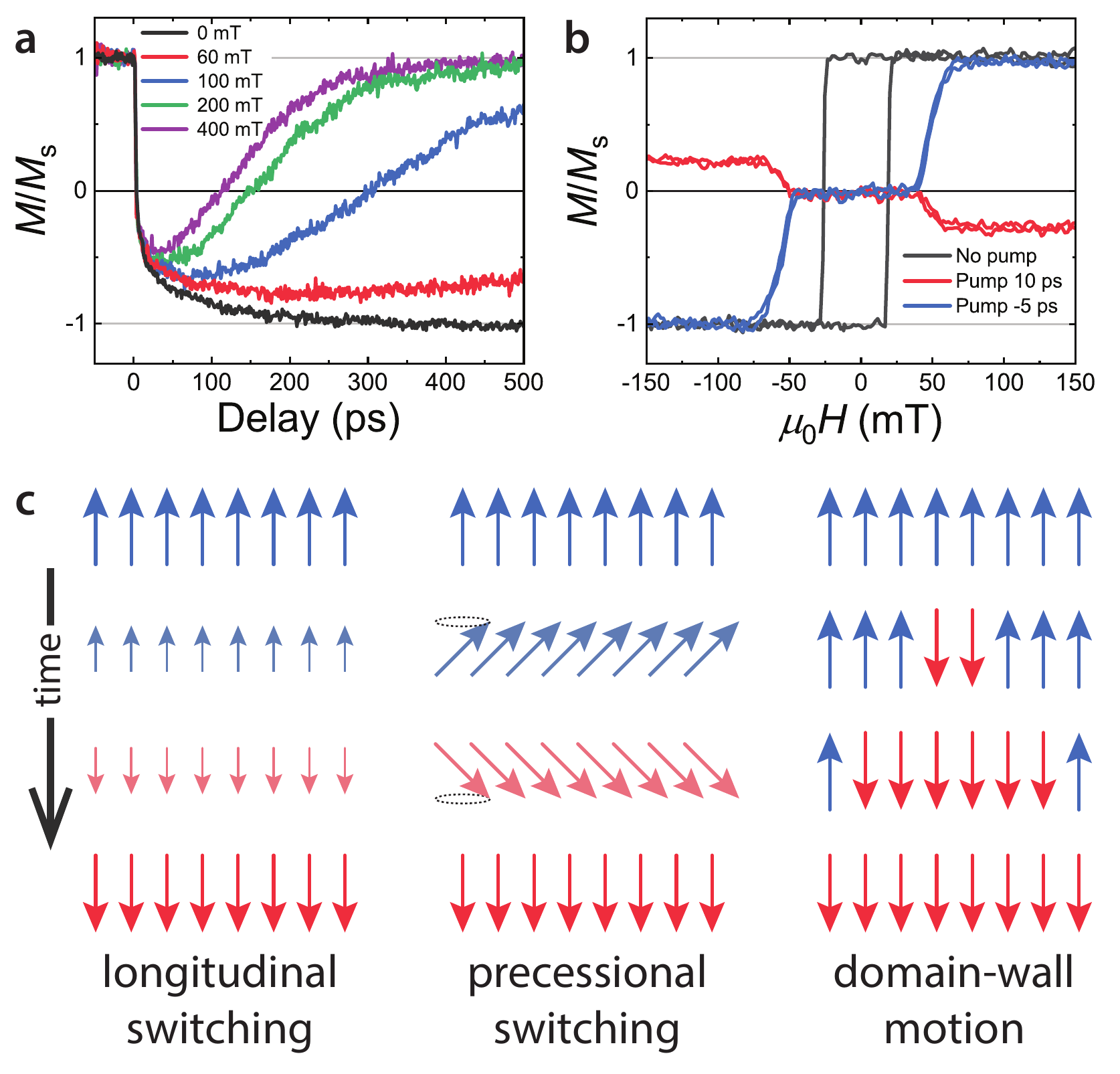}
	\caption{a) Time-resolved measurements of AOS with an applied magnetic field. The constant field is applied in the positive $z$-direction. b) Hysteresis loops with and without pump. The black line shows the hysteresis loop without pump, the red and blue lines show the hysteresis loops with pump and the delay fixed at \SI{-5}{\pico\second} and \SI{10}{\pico\second}, respectively. c) Schematic depiction of the three switching mechanisms examined in this paper: longitudinal switching, precessional switching and switching via domain-wall motion. }
	\label{fig:figure2CoGd}
\end{figure}

After having resolved the field-free switching, we turn towards the influence of a constant magnetic field, applied in the out-of-plane direction, on the switching dynamics. For strong enough fields, this will make sure that the magnetization switches back to the original magnetization direction between two successive pump pulses. For this reason this method is often used in pump-probe experiments. The influence of the field on the switching dynamics up to \SI{500}{\pico\second} can be seen in \cref{fig:figure2CoGd}a. We indeed observe the magnetization reversing back to the original direction due to the applied field, for fields larger than \SI{100}{\milli\tesla} even within the measured \SI{500}{\pico\second}. This answers the second question we posed: even at a sub-ns timescale the dynamics are indeed dependent on the magnetic field. For fields of several hundred \si{\milli\tesla} we observe a deviation with the zero-field measurement already after tens of \si{\pico\second}, while within hundreds of \si{\pico\second} the magnetization switches back. Such behaviour should be considered  surprising, realizing that the AOS process is governed by exchange fields of the order of \SI{1000}{\tesla}, more than three orders of magnitude larger than the applied field. The out-of-plane hysteresis loops visible in \cref{fig:figure2CoGd}b give some more insight in the fields and timescales related to the magnetization reversal due to the applied magnetic field. In black a regular hysteresis loop is visible, measured without the pump pulses present - so no AOS happens in this case. The loops in blue and red are measured with pump pulses present (so AOS does happen in this case) at two different delays between the pump and probe. In blue the hysteresis loop is visible with the delay fixed at \SI{10}{\pico\second}, at which the magnetization is (partially) switched. This can be seen in the hysteresis loop and also in \cref{fig:figure1CoGd}b, both showing $M/M_\mathrm{s}\approx 0.25$ after \SI{10}{\pico\second}. In these hysteresis loops we also see that fields up to \SI{150}{\milli\tesla} do not influence the dynamics on this timescale. In the red loop the delay is fixed at \SI{-5}{\pico\second}, which corresponds to the probe pulse arriving about \SI{10}{\micro\second} after the pump pulse. For small fields the signal is zero, as the magnetization toggles between up and down, resulting in no net signal. Above about \SI{55}{\milli\tesla} the magnetization is at the same level as without pump, showing that the magnetization has fully returned to the initial direction after the switch. For this reason, a field of \SI{55}{\milli\tesla} is the minimum field for which we are able to measure the dynamics of AOS at \SI{100}{\kilo\hertz}.

The question arises as to what the driving force behind this reversal and successive return at suprisingly short timescales is. We will discuss three mechanisms, as schematically depicted in \cref{fig:figure2CoGd}c. First we address longitudinal switching, the driving mechanism behind AOS. We will model this process using a modified microscopic three-temperature model (M3TM) \cite{Beens_PRB_2019}, as will be discussed in the next section. We conjecture that for timescales larger than \SI{20}{\pico\second} precessional switching, describing the switching due to applied torques on the magnetization, may play a role. This will be described using a Landau-Lifshitz-Bloch (LLB) model, which will be introduced in \cref{sec:LLB}. Finally, we will discuss the role of field-driven domain-wall motion, which occurs on even longer timescales and smaller fields, and whereby the switched domain will collapse even if the two other mechanisms are too weak.

\subsection{Longitudinal switching (M3TM)}
The first mechanism we discuss is longitudinal switching, the process that is responsible for AOS in the first place. Typically, this works on very short timescales ($<$\SI{10}{\pico\second}), but as can be seen in \cref{fig:figure2CoGd}a a deviation between the experimental traces at different magnetic fields occurs on short timescales already. The relevant field strengths in the switching process are the exchange fields, orders of magnitude larger than the fields that are applied in our experiment. The question is therefore whether the relatively small applied fields can significantly influence the switching process and could drive the backswitch via longitudinal switching, thereby explaining the dependence on the applied field in our experiments.

\begin{figure}[tb]
	\centering
	\includegraphics[width=\columnwidth]{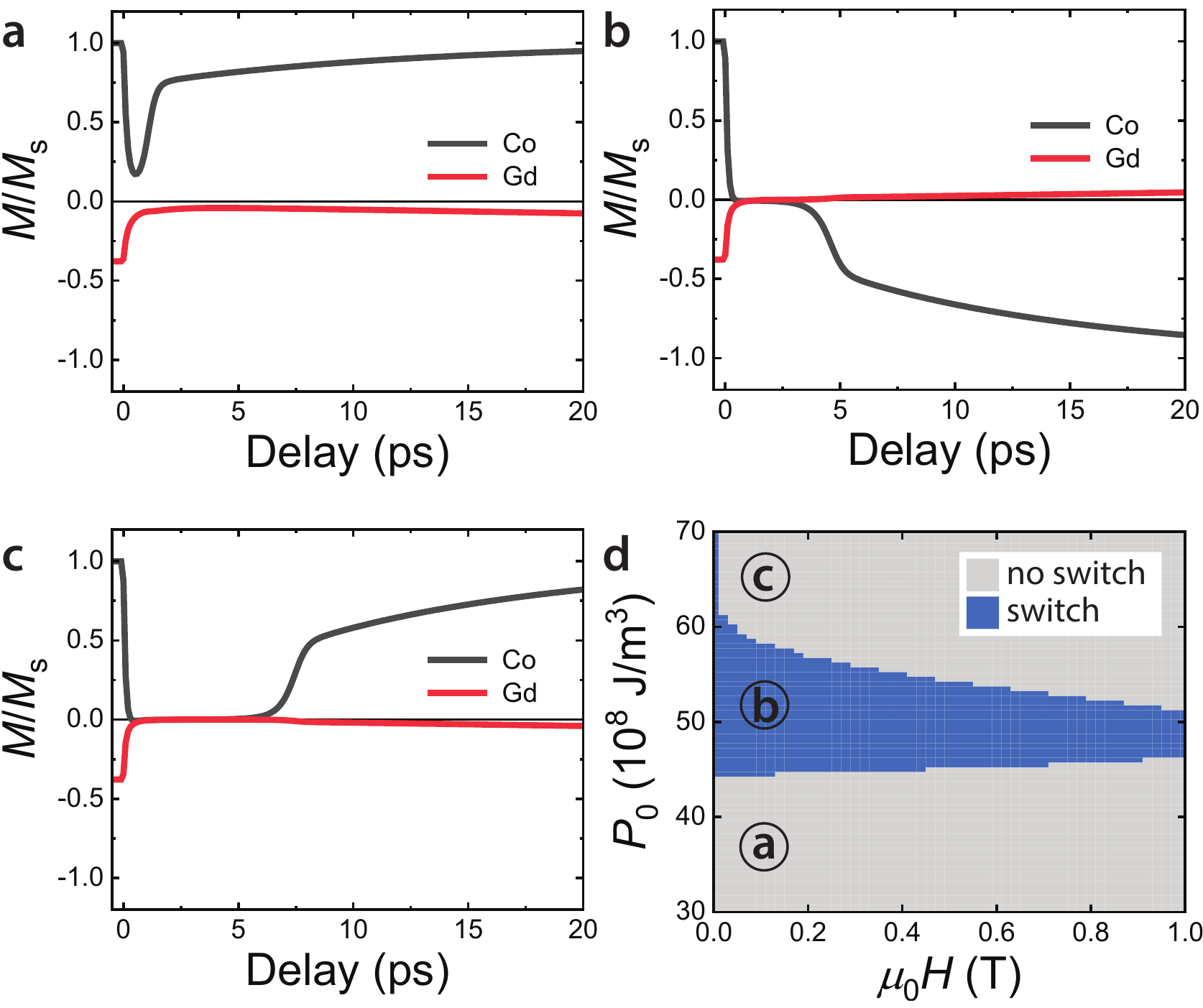}
	\caption{M3TM calculations. a) $m_z$ after laser pulse excitation with a fluence below the threshold fluence. b) $m_z$ after excitation with a laser pulse above the threshold fluence. c) $m_z$ after excitation with a laser pulse above the threshold fluence and a magnetic field in the positive $z$-direction strong enough to result in no switch. d) Phase plot of the magnetization direction after \SI{100}{\pico\second} as a function of applied field and the laser fluence. The three regions correspond to the traces shown in subfigures a, b and c.}
	\label{fig:figure3CoGd}
\end{figure}

To simulate the switching process we use a layered-M3TM \cite{Koopmans_NatMat_2009,Beens_PRB_2019}, which models each atomic layer as a separate system, interacting with neighboring layers via exchange scattering. In Supplementary Note III.A \cite{supplementary} more details about this model can be found. In brief, we use five atomic layers of Co and three atomic layers of Gd, roughly corresponding to \SI{1}{\nano\meter} of Co and \SI{1}{\nano\meter} of Gd. The ground state magnetization profile of the Gd decays exponentially, and adding more Gd layers would not significantly change the resulsts. A Gaussian laser pulse is used to excite the system, after which the magnetization dynamics are calculated in a layer-resolved fashion. In \cref{fig:figure3CoGd}a the result of such a simulation is depicted for a relatively low laser energy, with the black and red line indicating the average Co and Gd magnetization, respectively. In this case, the laser fluence is not high enough for the electron temperature to reach the Curie temperature. Thus, after partial demagnetization the magnetization of both layers returns back to the initial direction, and there is no AOS. Above a certain switching threshold AOS is observed, as shown in \cref{fig:figure3CoGd}b. A transient ferromagnetic state at a very low magnitude is visible as a plateau for several \si{\pico\second}, after which the Gd magnetization crosses zero as well, which creates a switched state. In both these cases (\cref{fig:figure3CoGd}a and \cref{fig:figure3CoGd}b) no applied field is used. When introducing an applied field, aligned with the original Co magnetization, a third possibility appears. For the right combination of fluence and field strength, the magnetization is almost fully quenched, but returns to its original direction instead of switching, which can be seen in \cref{fig:figure3CoGd}c. This applied field-driven backswitch is a consequence of the long duration of the plateau that lasts for several \si{\pico\second}. In this state with almost fully quenched magnetization, a relatively small magnetic field is enough to prevent the switch and cause the magnetization to go back to its original direction. In \cref{fig:figure3CoGd}d a phase plot is visible where we show how the combination of laser fluence and applied magnetic field influences the switching behavior. The switched and non-switched regions in the phase plot are determined by tracking the direction of the Co magnetization after \SI{100}{\pico\second} and comparing that with the original direction. For laser fluences below the threshold fluence no switching happens, as visible in the bottom grey area (a). For increasing magnetic fields, we see a slight increase in the threshold fluence. Above this area the region where switching occurs is visible (b). For zero field these are the only two possibilities, but for a finite field a third region (c) exists, which corresponds to the magnetization returning to its original direction. From the phase plot it is clear that this effect is a combination of high fluence and large magnetic field. This second threshold for backswitching strongly depends on the applied magnetic field, with a higher fluence requiring a smaller magnetic field to achieve this. This can be understood from the strong reduction of effective exchange field in the strongly quenched magnetic state, resulting in applied fields and exchange fields of similar magnitudes.

Although this behavior looks similar to what we see in our experiment, where an external field can result in the magnetization returning to the original direction, the mechanism at play does not appear to be the same. The pronounced plateau that can be seen in these calculations is not visible in our experiments, and we verified that this plateau is the crucial driving force for the backswitch mechanism in the layered-M3TM. We conjecture that the occurance of such a plateau is an artefact of the M3TM, as this model does not take into account thermal or spatial fluctuations. In models that do take into account these fluctuations, such as atomistic LLG calculations, this plateau is not visible, more in line with the experiment. As a consequence, we expect the strong field dependence not to be visible in such models. We do stress that there might be a mechanism of longitudinal switching playing a role in the switching back, although one that is not captured by our simple model.

\subsection{Precessional switching (LLB)}
\label{sec:LLB}
The second mechanism that we discuss is precessional switching, which describes the behavior of the magnetization under the influence of torques applied to it, in our case due to a combination of the externally applied field and the local anisotropy fields. To describe this behavior we use an LLB model \cite{Atxitia_JPD_2016}, where we explicitly take into account the change in magnitude of the magnetization because of the AOS. This results in the following equation for the transverse dynamics:

\begin{equation}
\frac{\mathrm{d}\mathbf{m}}{\mathrm{d}t} = -\frac{\gamma \mu_0}{1+\alpha^2} \left( \mathbf{m}\times\mathbf{H}_\mathrm{eff} + \alpha \chi(t) \mathbf{m}\times\mathbf{m}\times\mathbf{H}_\mathrm{eff} \right), 
\label{eqn:llb}
\end{equation}
% Equation divided by chi(t) on both sides to make sure it fits on one line
with $\mathbf{m}$ the normalized magnetization, $\gamma$ the gyromagnetic ratio, $\mu_0$ the vacuum permeability, $\alpha$ the Gilbert damping constant and $\mathbf{H}$ the effective field\textemdash comprised of the external field and the anisotropy field. Furthermore, $\chi(t)$ is the time-dependent magnitude of the magnetization given by a double exponential fit to the data from \cref{fig:figure1CoGd}b, which starts at a positive value and becomes negative after the switch. Thus, $|\textbf{M}(t)| = \chi(t) M_\mathrm{s}$, where we assume that $\chi(t)$ is not affected by the magnetic field. In Supplementary Note III.B \cite{supplementary} a more detailed explanation of the model is given. 

In our simulations we define the $z$-axis as the OOP axis. The initial magnetization is in the $z$-direction and the magnetic field in the $xz$-plane, making an angle $\theta$ with the $z$-axis. This is done to incorporate the effect of a possible small misalignment of the magnetic field with the easy axis of the sample in the experiments, as well as the fact that due to thermal fluctuations on a microscopic scale there will be a finite angle between the local magnetization and the easy axis. Rather than describing an inhomogeneous system with a spatially (and temporally) fluctuating local magnetization, we limit ourselves to a single macrospin with a fixed orientation with respect to the applied field to capture the basic physics. The parameters used in these calculations are realistic values for a Co/Gd bilayer system and can be found in Supplemental Note III.B.

\begin{figure}[tb]
	\centering
	\includegraphics[width=\columnwidth]{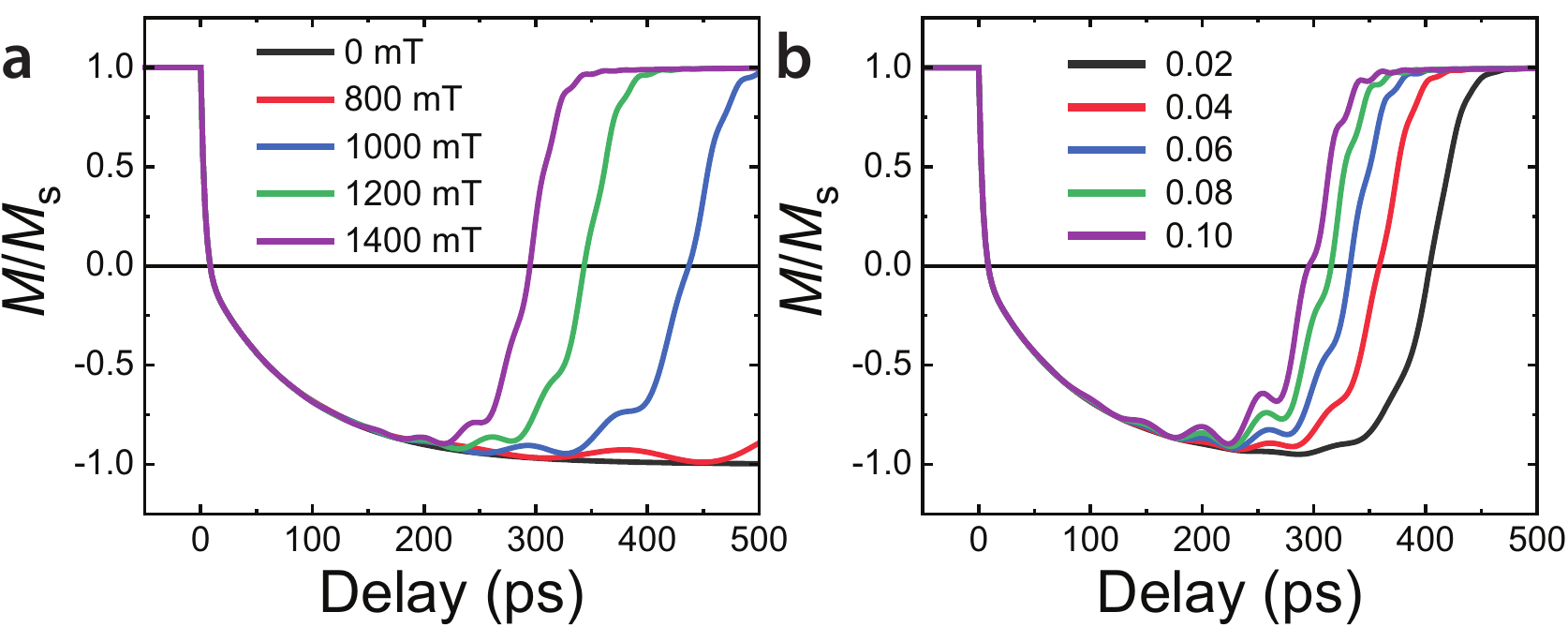}
	\caption{LLB calculations. a) $z$-magnetization as a function of delay for various magnetic fields, pointing in the positive $z$-direction. The angle between the applied field and the $z$-axis is $\theta = \SI{0.05}{\radian}$. b) $z$-magnetization as a function of delay for various values of $\theta$. The applied field is $\mu_0 H = \SI{1200}{\milli\tesla}$.}
	\label{fig:figure4CoGd}
\end{figure}

We first let the magnetization relax to an equilibrium state, after which at $t=0$ the magnetization starts to switch to the opposite direction via longitudinal dynamics described by $\chi(t)$. In \cref{fig:figure4CoGd}a the magnetization in the $z$-direction from these calculations is shown for various magnetic fields, assuming $\theta = \SI{0.05}{\radian}$. For the first \SI{200}{\pico\second} no significant influence of the applied field is visible. In all cases there is a demagnetization, followed by a reversal of the magnetization direction and a remagnetization in the opposite state. On longer timescales the magnetization reverses to the initial direction - more quickly for the higher applied fields. Qualitatively, this corresponds well with the experiments (\cref{fig:figure2CoGd}a), as the magnetization reverses back to the original direction within \SI{500}{\pico\second}. However, several differences are noteworthy. First, the fields needed to establish a sub-\si{\nano\second} backswitch in the calculations are a factor two to three higher than the fields in the experiment. This might be explained by our neglect of thermal fluctuations. Although in the model the angle $\theta$ between easy axis and magnetic field direction plays a similar role, this should be considered a severe approximation. Furthermore, we notice that in our calculations up to \SI{200}{\pico\second} no field dependence is seen, in contrast to our experiments. Moreover, the reversal happens faster in the calculations than in the experiments. Apart from the already mentioned thermal fluctuations, another factor that might explain these differences is the assumption of a single macrospin, instead of the two coupled subsystems that are present in the bilayer. To explore other reasons for the quantitative mismatch between simulations and experiments, we have looked into the influence of the damping $\alpha$, the effective anisotropy $K_\mathrm{eff}$ and the saturation magnetization $M_\mathrm{s}$ as well, but found no qualitative difference in the switching process. Details of these calculations can be found in Supplementary Note IV \cite{supplementary}. 

Note that the oscillations that we observe in \cref{fig:figure4CoGd} are a trivial consequence of our macro-spin approach. They emerge as a result of the angle between the easy axis and $\textbf{H}$, mainly visible for bigger angles as can be seen in \cref{fig:figure4CoGd}b. Therefore we do not expect to these oscillations to occur in experiments. 

To conclude, the dynamics of the backswitch within \SI{500}{\pico\second} are thus qualitatively well explained by our simple LLB model, but some detailed features cannot be fully reproduced, mainly with regards to the behavior within \SI{200}{\pico\second}. While this is partly due to simplifications made in our model, it is possible that other effects might play a role as well. For example, the longitudinal time-dependence of the magnetization is assumed to be independent of the magnetic field.  A more realistic approach that takes into account the field dependence of $\chi(t)$ may be needed for a better agreement with the experiments.

\subsection{Domain-wall motion}
A third switching mechanism is domain-wall motion. If the applied field is in the same direction as the magnetization outside of the switched region, it will cause the switched domain to shrink and eventually collapse. Assuming a switched domain with a radius of \SI{15}{\micro\meter} and a domain-wall velocity of \SI{200}{\meter\per\second} at \SI{60}{\milli\tesla}\cite{Pham_EPL_2016} we find that it takes \SI{75}{\nano\second} for the domain to disappear, and thus that within \SI{500}{\pico\second} the DW travels \SI{0.1}{\micro\meter}. Relating this to the measurements from \cref{fig:figure2CoGd}a, it is safe to conclude that the dynamics in this measurement are unaffected by the influence of DW motion via the edge of the switched domain, as the domains have a diameter of $\sim \SI{30}{\micro\meter}$. However, this assumes a single nucleation site within the switched region. If more backswitched domains are formed, that will increase the effect of DW motion on the total switched area. However, only for a large number of nucleation sites this is expected to be significant, as these individual DWs still travel only \SI{0.1}{\micro\meter} in \SI{500}{\pico\second}.

\begin{figure}[tb]
	\centering
	\includegraphics[width=0.7\textwidth]{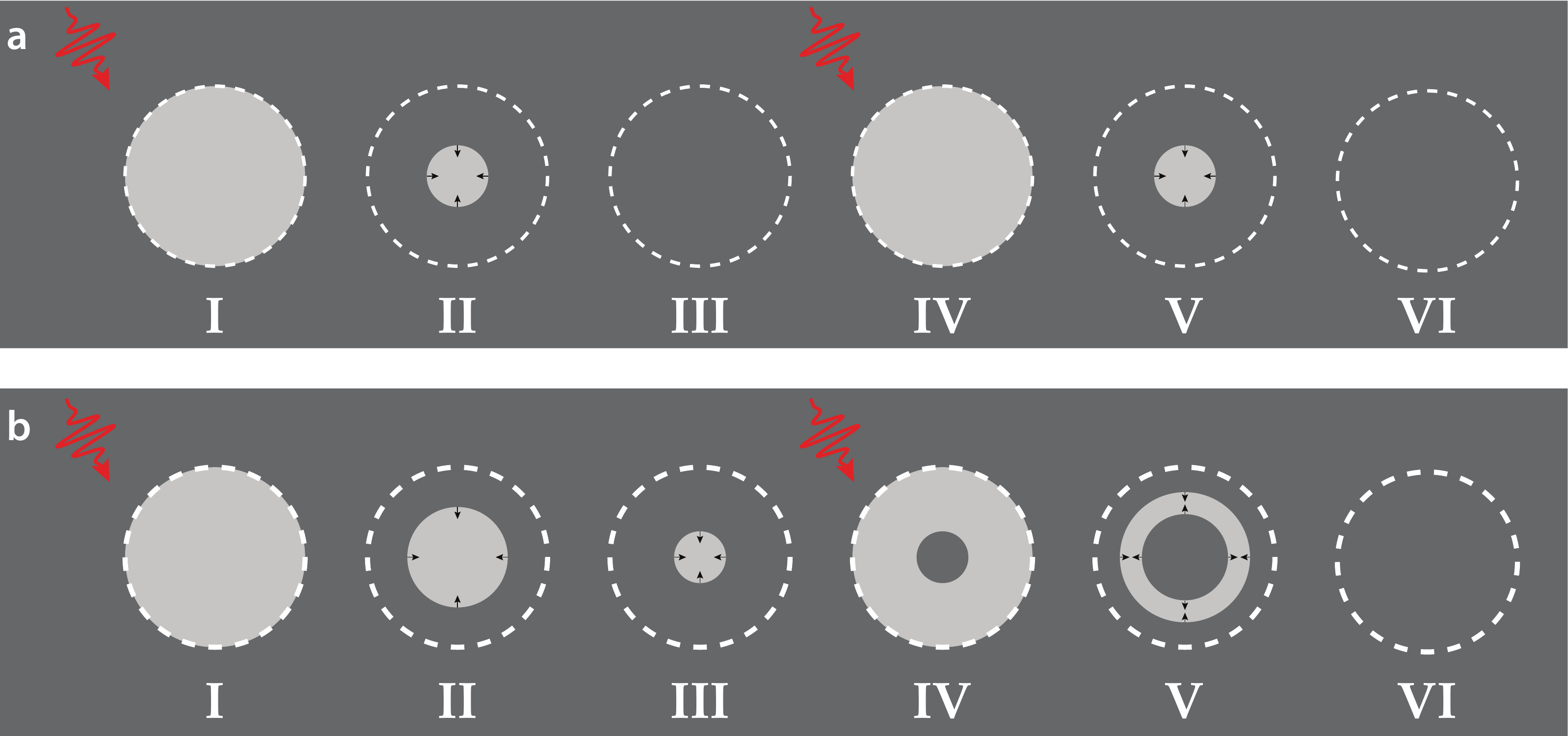}
	\caption{Schematic depiction of the cycles that can appear in the switching process due to domain-wall motion and AOS, with time going from left to right. The black arrows indicate the direction of DW-motion due to the applied field. a) Cycle for an applied field strong enough to fully reverse the magnetization between two pulses. b) Cycle for a smaller field, resulting in a longer cycle.}
	\label{fig:figure5CoGd}
\end{figure}

Although we excluded DW motion to be of relevance for the dynamics within \SI{500}{\pico\second}, it probably does play a role in ensuring the magnetization reverses to its original direction between two successive pulses. The DW velocities mentioned above are certainly high enough to achieve a collapse within \SI{10}{\micro\second} (the time between two successive pulses). In Figure \ref{fig:figure5CoGd}a this process is shown, with the dark regions corresponding to the up magnetization and the light regions corresponding to the down magnetization. Time goes from left to right, with first a switched domain (I), then the domain shrinking due to domain-wall motion (II) and fully disappearing (III) before the next pulse arrives (IV). 

For this process a certain field is needed for which the magnetization is fully reversed before the next pump pulse arrives. From \cref{fig:figure2CoGd}b we estimate that the minimum applied field is approximately \SI{60}{\milli\tesla}, corresponding to a DW velocity of \SI{1.5}{\meter\per\second}. If the field is smaller than that, a more intricate progression may appear. This is represented in \cref{fig:figure5CoGd}b. Again we see a switched domain (I), then the domain shrinking due to the applied field (II \& III). However, now the domain is not fully reversed, and a ring pattern is expected to appear after the next pulse (IV). This ring pattern will again shrink, now from both sides (V). Due to this the domain can be fully reversed before the next pulse arrives (VI), and the cycle starts from the beginning. Probing a corresponding doubling of the periodicity would provide direct evidence of this DW motion driven dynamics, but is beyond the scope of our present paper.

\section{Conclusions}
To conclude, in this paper we showed field-free measurements of the switching dynamics in Co/Gd bilayers, by using the toggle-switching behavior of HI-AOS and measuring every other pulse. We confirm that AOS in Co/Gd happens at the same ultrafast rate as in the alloys CoGd and GdFeCo. When applying an external field, we observe a strong field dependence within \SI{500}{\pico\second}, visible as a switch back to the initial magnetization direction. Discussing various switching mechanisms (longitudinal switching, precessional switching, DW motion) we conclude that the fast backswitch is dominated by precessional switching, but that this does not fully explain the observed behavior, especially on timescales below \SI{50}{\pico\second}. Furthermore, we conjecture that at smaller fields a collapse of the switched domain will be driven by DW motion. We predict the possibility to observe a doubling of the periodicity of the dynamics in this regime. 

Finally, we expect our work to trigger further research aimed at more quantitative explanation of the magnetic field dependent behavior. Independent of the underlying mechanism, it will be important to keep in mind all these effects when analyzing time-resolved AOS experiments in the presence of an applied field.

\begin{acknowledgments}
We thank Gregory Malinowski, Jon Gorchon and Stephane Mangin for their help and fruitful discussions during the visit of MJGP and we thank Maarten Beens for the help with the layered-M3TM. The work is part of the research programme of the Foundation for Fundamental Research on Matter (FOM), which is part of the Netherlands Organisation for Scientific Research (NWO).
\end{acknowledgments}

\bibliography{library}

\end{document}

\title{Supplementary Material: Influence of magnetic fields on switching dynamics in Co/Gd bilayers}

\author{M.J.G. Peeters}
\email[E-mail: ]{m.j.g.peeters@tue.nl}
\affiliation{Department of Applied Physics, Eindhoven University of Technology, PO Box 513, 5600 MB Eindhoven, The Netherlands}

\author{Y.M. van Ballegooie}
\affiliation{Department of Applied Physics, Eindhoven University of Technology, PO Box 513, 5600 MB Eindhoven, The Netherlands}

\author{B. Koopmans}
\affiliation{Department of Applied Physics, Eindhoven University of Technology, PO Box 513, 5600 MB Eindhoven, The Netherlands}

\date{\today}

\maketitle

\section{Setup}
\begin{figure}[tb]
	\centering
	\includegraphics[width=0.6\textwidth]{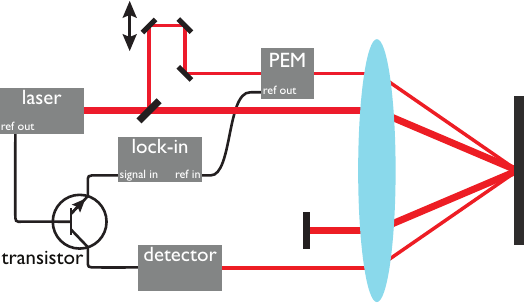}
	\caption{Schematic depiction of the TR-MOKE setup, including a photo-elastic modulator as a modulation technique and a transistor to achieve half-pulse measurements. Not all optical components are shown in this schematic.}
	\label{fig:suppcogd_setup}
\end{figure}

In \cref{fig:suppcogd_setup} the setup is schematically depicted. The laser output of the Spectra-Physics Spirit-NOPA system has a central wavelength of \SI{700}{\nano\meter} and a repetition rate of \SI{100}{\kilo\hertz}. The pulse length is \SI{35}{\femto\second} at the laser output and $\sim$\SI{100}{\femto\second} at sample position. The laser output is split in high intensity pump pulses and low intensity probe pulses by a 90:10 beamsplitter. The pump pulses are sent to the sample to initiate the switching. These pump pulses are focused to a spot size with a FWHM of $\sim$\SI{80}{\micro\meter} at sample position, compared with $\sim$\SI{30}{\micro\meter} for the probe pulses. This ensures that the probed region is fully switched. The probe pulses are sent through a delay line with which the relative delay between pump and probe pulses at sample position can be changed. This allows us to do time-dependent measurements. A polarizer (not pictured) and a photo-elastic modulator are included for polarization modulation, at a frequency of \SI{50}{\kilo\hertz}. The reflection of the probe pulses is sent through a second polarizer (the analyzer, also not pictured) and collected by the detector, after which it is sent through the transistor and to the input of the lock-in amplifier. The base of the transistor is connected to a function generator, which generates a square wave with half the frequency of the repetition rate of the laser. A high signal to the base of the transistor will block the signal to the lock-in, while a low signal will let it pass. The function generator is synchronized with the reference output of the laser. This allows us to measure only the up-down or down-up switches by blocking every other pulse. 

%\subsection{Kerr microscope}
%\textcolor{red}{Checks for good measurements in Co/Gd (fluence dependence etc. in Kerr microscope).}
%
%\begin{figure}[tb]
%	\centering
%	\includegraphics[width=0.6\textwidth]{SuppCoGd_Kerr-01.eps}
%	\caption{Fluence dependence.}
%	\label{fig:suppcogd_kerr}
%\end{figure}

\section{Validity check transistor measurement}
\begin{figure}[tb]
	\centering
	\includegraphics[width=0.5\textwidth]{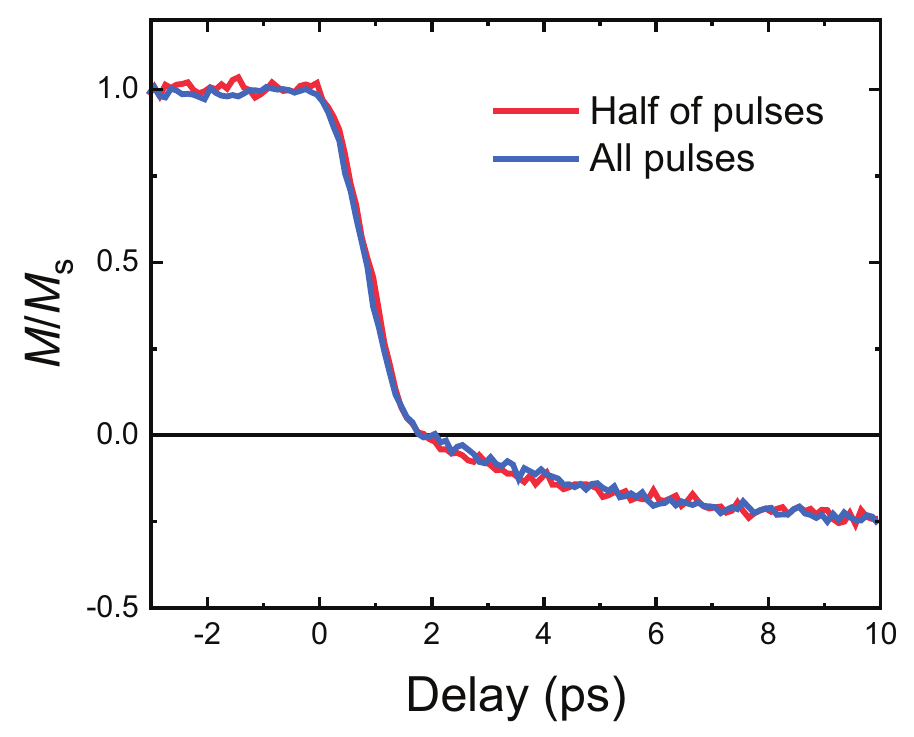}
	\caption{Comparison between measurement with all pulses and with only half of the pulses, measured with a field of \SI{55}{\milli\tesla}.}
	\label{fig:suppcogd_allpulses}
\end{figure}

To confirm that using the transistor to measure only half of the pulses does not introduce any artifacts in our measurements we compare a measurement with transistor (thus measuring only half of the pulses) with a measurement without transistor (measuring all pulses). A constant applied field of \SI{55}{\milli\tesla} is used in both measurements, as for lower fields the magnetization is not fully reversed in the time between two pulses (see Figure 2b). The measurements presented in \cref{fig:suppcogd_allpulses} indeed show no difference, thus showing the validity of this method. 

%\subsection{Hysteresis with pump}
%Hysteresis loop with pump (negative/positive time delay), if not put in paper
%
%\begin{figure}[tb]
%	\centering
%	\includegraphics[width=0.6\textwidth]{SuppCoGd_Hyst-01.eps}
%	\caption{Hysteresis loop with pump, positive and negative time delay, and without pump.}
%	\label{fig:suppcogd_hyst}
%\end{figure}

\section{Model details}
In this section we describe the details of the two models used in this article, the layered-M3TM describing longitudinal switching and the LLB model describing precessional switching. 

\subsection{Layered-M3TM}
The layered-M3TM is an extension of the general M3TM where the layered nature of the modeled material is explicitly taken into account, which we use to model all-optical switching. First we give a very brief overview of the most important equations for the general M3TM (for a more detailed description, see \cite{Koopmans_NatMat_2009}), followed by the characteristics of the layered-M3TM (for a more detailed description, see \cite{Beens_PRB_2019}). 

In the general M3TM the ferromagnetic system is described by three subsystems: the electron system, the phonon system and the spin system. The electron and phonon subsystems are described by the equations
\begin{align}
\gamma T_\mathrm{e} \frac{\mathrm{d}T_\mathrm{e}}{\mathrm{d}t} &= g_\mathrm{ep}(T_\mathrm{p}-T_\mathrm{e}) + P(t), \label{eqn:L-M3TM1}\\
C_\mathrm{p}\frac{\mathrm{d}T_\mathrm{p}}{\mathrm{d}t} &= g_\mathrm{ep}(T_\mathrm{e}-T_\mathrm{p}), \label{eqn:L-M3TM2}
\end{align}
where $T_\mathrm{e}$ and $T_\mathrm{p}$ are the electron and phonon temperatures, $C_\mathrm{p}$ is the phonon heat capacity, $g_\mathrm{ep}$ is the electron-phonon interaction constant and $P(t)$ describes the laser pulse excitation that heats the electron system. 

For the spin subsystem a Weiss mean field model is used, where we assume that it can be described as a $S = 1/2$ system. We assume a $S = 1/2$ system for both Co and Gd, although $S = 7/2$ is possible for Gd as well \cite{Beens_PRB_2019}. An Elliot-Yafet scattering mechanism is used to implement the angular momentum dissipation. Using Fermi's Golden Rule we can then describe the relation between the normalized magnetization $m$, $T_\mathrm{e}$ and $T_\mathrm{p}$ via 
\begin{equation}
\frac{\mathrm{d}m}{\mathrm{d}t} = R \frac{T_\mathrm{p}}{T_\mathrm{C}}m\left[1-m\coth{\left(\frac{T_\mathrm{C}}{T_\mathrm{e}}m\right)}\right],
\label{eqn:m3tmregular}
\end{equation}
with $R$ the demagnetization rate and $T_\mathrm{C}$ the Curie temperature. The splitting of energy levels between up and down spins, $\Delta_\mathrm{ex}$, is given by (for $S = 1/2$)
\begin{equation}
\Delta_\mathrm{ex} = 2k_\mathrm{B}T_\mathrm{C},
\end{equation}
with $k_\mathrm{B}$ the Boltzmann constant. The equilibrium magnetization for a given ambient temperature $T_\mathrm{amb}$ is then given by 
\begin{equation}
m = \tanh{\frac{\Delta_\mathrm{ex}}{2k_\mathrm{B}T_\mathrm{amb}}}.
\end{equation}

In the layered-M3TM the monolayers of the magnetic material are treated as separate systems, interacting with neighboring layers via the exchange interaction. For each of these layers $i$ the equilibrium magnetization $m_i$ is given by
\begin{equation}
m_i = \tanh{\frac{\Delta_{\mathrm{ex},i}}{2k_\mathrm{B}T_\mathrm{amb}}},
\end{equation}
with $\Delta_{\mathrm{ex},i}$ given by 
\begin{equation}
\Delta_{\mathrm{ex},i} = 2 k_\mathrm{B}\left(\frac{3T_{\mathrm{C},i-1}m_{i-1} + 6T_{\mathrm{C},i}m_{i} + 3T_{\mathrm{C},i+1}m_{i+1}}{12}\right).
\end{equation}
Here we have assumed an fcc lattice for the magnetic structure, with 6 nearest neighbors in the same layer and 3 in the neighboring layers. The differential equation for the magnetization in the layered-M3TM for a given layer is a combination of the dynamics within the layer and the exchange scattering with the neighboring layers,
\begin{equation}
\frac{\mathrm{d}m_i}{\mathrm{d}t} = \left.\frac{\mathrm{d}m_i}{\mathrm{d}t}\right|_i + \left.\frac{\mathrm{d}m_i}{\mathrm{d}t}\right|_{\mathrm{ex},i,i+1} + \left.\frac{\mathrm{d}m_i}{\mathrm{d}t}\right|_{\mathrm{ex},i,i-1}.
\label{eqn:L-M3TM3}
\end{equation}
The first term in this equation is the equivalent of \cref{eqn:m3tmregular} for a single layer, while the second and third term describe the exchange scattering with the neighboring layers. In Ref. \cite{Beens_PRB_2019} the full version of these exchange scattering terms can be found. For the exchange interaction between Co and Gd we use $J_\mathrm{ex}/k_\mathrm{B} = \SI{-1000}{\kelvin}$, with $J_\mathrm{ex}$ equal to $\Delta_\mathrm{ex}$ for $m = 1$. This value was taken from \cite{Lalieu_thesis}.

The electron and phonon systems are again described by global temperatures, $T_\mathrm{e}$ and $T_\mathrm{p}$. These temperatures are assumed to be in equilibrium throughout the different layers, and are described by \cref{eqn:L-M3TM1} and \cref{eqn:L-M3TM2}. Adding heat diffusion from the phonon system to the surroundings to \cref{eqn:L-M3TM2}, these two equations combined with \cref{eqn:L-M3TM3} form the layered-M3TM. For Co and Gd we used material parameters as given in \cref{tab:m3tm}, where the shared parameters describing the electron and phonon system relate to Co. 

\begin{table}[]
\caption{Table with values and descriptions of parameters used in the layered-M3TM. Parameters are taken from Ref. \cite{Lalieu_thesis}.}
\label{tab:m3tm}
\begin{tabular}{p{3cm}p{2cm}p{2cm}l}
\textbf{Parameter}         & \multicolumn{2}{l}{\textbf{Value}}                                                          & \textbf{Description}                          \\ \hline
$\gamma$          & \multicolumn{2}{l}{\SI{2000}{\joule\per\meter\cubed\per\kelvin\squared}}           & electron heat capacity constant      \\ 
$C_\mathrm{p}$    & \multicolumn{2}{l}{\SI{4e6}{\joule\per\meter\cubed\per\kelvin}}                    & phonon heat capacity                 \\ 
$g_\mathrm{ep}$   & \multicolumn{2}{l}{\SI{4.05e6}{\joule\per\meter\cubed\per\kelvin\per\pico\second}} & electron-phonon interaction constant \\ 
$J_\mathrm{ex}/k_\mathrm{B} $          & \multicolumn{2}{l}{\SI{-1000}{\kelvin}}           & Co-Gd exchange      \\ 
$\tau$            & \multicolumn{2}{l}{\SI{10}{\pico\second}}                                          & heat diffusion time                  \\ 
$\sigma$          & \multicolumn{2}{l}{\SI{0.05}{\pico\second}}                                        & laser pulse duration                 \\ 
$T_\mathrm{amb}$  & \multicolumn{2}{l}{\SI{295}{\kelvin}}                                              & ambient temperature                  \\ 
                  & \textbf{Co}                                       & \textbf{Gd}                                      &                                      \\
$T_\mathrm{C}$    & \SI{1388}{\kelvin}                       & \SI{292}{\kelvin}                       & Curie temperature                    \\
$a_\mathrm{sf}$   & 0.15                                     & 0.08                                    & spin-flip probability                \\
$\mu_\mathrm{at}$ & 1.72                                     & 7.55                                    & atomic magnetic moment               \\
$r_\mathrm{at}$   & \SI{1.35}{\angstrom}                     & \SI{1.86}{\angstrom}                    & atomic radius                        \\
$E_\mathrm{D}$    & 400                                      & 400                                     & Debye energy                         \\
$D_\mathrm{F}$     & \SI{3}{\per\electronvolt}                & \SI{3}{\per\electronvolt}               & density of states                    \\
$D_\mathrm{S}$    & 1.72                                     & 7.55                                    & number of spins per atom            
\end{tabular}
\end{table}

\subsection{LLB model}
As described in the main text, our implementation of the LLB model is given by 
\begin{equation}
\frac{\mathrm{d}\mathbf{m}}{\mathrm{d}t} = -\frac{\gamma \mu_0}{1+\alpha^2} \left( \mathbf{m}\times\mathbf{H}_\mathrm{eff} + \alpha \chi(t) \mathbf{m}\times\mathbf{m}\times\mathbf{H}_\mathrm{eff} \right).
\label{eqn:llb_supp}
\end{equation}
% Equation divided by chi(t) on both sides to make sure the equation fits on one line in main text
The (non-normalized) magnetization $\textbf{M}(t)$ is related to $\textbf{m}(t)$ and $\chi(t)$ via 
\begin{equation}
\textbf{M}(t) = \textbf{m}(t)\chi(t)M_\mathrm{s}.
\end{equation}
$\chi(t)$ is approximated by a double exponential, and describes the time dependence of the longitudinal component of the magnetization:
\begin{equation}
\chi(t) = 1 + H(t)\left[e^{-\frac{t}{\tau_1}}+e^{-\frac{t}{\tau_2}}-2\right],
\end{equation}
with $H(t)$ the Heaviside step function and $\tau_1 = \SI{3.6}{\pico\second}$ and $\tau_2 = \SI{86.7}{\pico\second}$, obtained by fitting the data from Figure 1b. The effective field in \cref{eqn:llb_supp} is a combination of the applied field and an anisotropy field, and is given by
\begin{equation}
\mathbf{H}_\mathrm{eff} = \mathbf{H}_\mathrm{app} + H_\mathrm{K}\hat{\mathbf{z}} = \mathbf{H}_\mathrm{app} + \frac{2 K_\mathrm{eff}}{\mu_0 M_\mathrm{s}}\chi(t)m_z(t)\hat{\mathbf{z}},
\end{equation}
with $\textbf{H}_\mathrm{app}$ the applied field, which is always in the $xz$-plane, and $m_z(t)$ the $z$-component of the normalized magnetization $\textbf{m}$. The meaning and values of the other parameters are given in \cref{tab:llb}. The effective anisotropy $K_\mathrm{eff}$ and the saturation magnetization $M_\mathrm{s}$ are taken from literature \cite{Cao_PRB_2020, Lalieu_PRB_2017}, while for the damping parameter $\alpha$ we assume that it is the same as the damping parameter in CoGd alloys \cite{Binder_PRB_2006}. In the next section we will see that the behaviour described in the main text is robust under changes to these parameters. 

\begin{table}[]
\caption{Table with values and descriptions of parameters used in the LLB model.}
\begin{tabular}{cll}
\textbf{Parameter}& \textbf{Value}                        & \textbf{Description}     \\
\hline
$\gamma$         & \SI{176}{\radian\per\second\per\tesla} & gyromagnetic ratio       \\
$\mu_0$          & \SI{4\pi e-7}{\henry\per\meter}        & vacuum permeability      \\
$\alpha$         & 0.15                                   & Gilbert damping constant \\
$K_\mathrm{eff}$ & \SI{0.12}{\mega\joule\per\meter\cubed} & effective anisotropy     \\
$M_\mathrm{s}$   & \SI{0.4}{\mega\ampere\per\meter}       & saturation magnetization \\
\end{tabular}
\label{tab:llb}
\end{table}

\section{Effect of $\alpha$, $M_\textrm{s}$ and $K_\textrm{eff}$ in LLB model}
\begin{figure}[tb]
	\centering
	\includegraphics[width=0.95\textwidth]{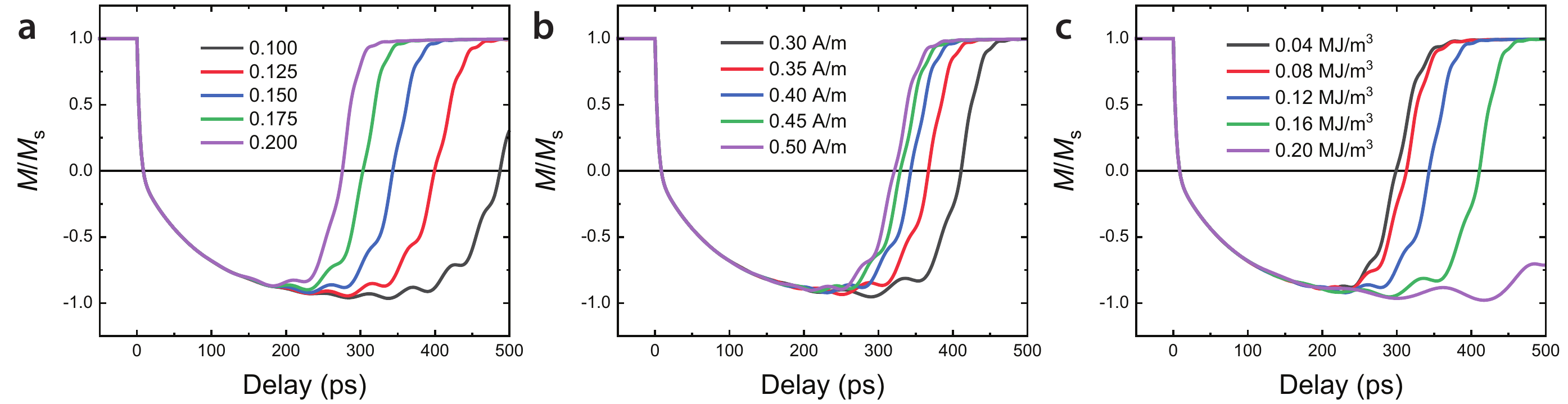}
	\caption{LLB calculations as a function of several parameters not discussed in the main text. All calculations shown here are done for $H_\mathrm{app} = \SI{1200}{\milli\tesla}$ and $\theta = 0.05$. a) Damping parameter $\alpha$. b) Saturation magnetization $M_\textrm{s}$. c) Effective anisotropy $K_\textrm{eff}$.}
	\label{fig:suppcogd_LLB}
\end{figure}

In \cref{fig:suppcogd_LLB} we show the influence of $\alpha$, $M_\textrm{s}$ and $K_\textrm{eff}$ on the magnetization reversal in the LLB model. As can be seen, qualitatively there are minimal differences when varying these parameters apart from the moment when the reversal occurs, which can be intuitively understood for all parameters. For the damping parameter $\alpha$ we observe earlier switching for higher damping, as is expected because the damping term in the LLB equation is the driving force behind the switching. For the saturation magnetization a higher $M_\mathrm{s}$ results in earlier switching, as this corresponds to a higher Zeeman energy gain. For the effective anisotropy, a lower anisotropy results in earlier switching as the anisotropy hinders the switching. 

\bibliography{../library}